\def\year{2026}\relax

\documentclass[letterpaper]{article} 
\usepackage{aaai25}  
\usepackage{times}  
\usepackage{helvet}  
\usepackage{courier}  
\usepackage[hyphens]{url}  
\usepackage{graphicx} 
\usepackage{natbib}  
\usepackage{caption} 
\DeclareCaptionStyle{ruled}{labelfont=normalfont,labelsep=colon,strut=off} 
\usepackage{placeins}
\usepackage{algorithm}
\usepackage{algorithmic}
\usepackage{booktabs}
\usepackage{amsmath}
\usepackage{multirow}
\usepackage{listings}
\usepackage{xcolor}
\newcommand{\answerYes}[1]{\textcolor{blue}{#1}} 
\newcommand{\answerNo}[1]{\textcolor{teal}{#1}} 
\newcommand{\answerNA}[1]{\textcolor{gray}{#1}}

\title{Auditing the Reliability of Multimodal Generative Search}

\author{
    Erfan Samieyan Sahneh\textsuperscript{\rm 1, \rm 2},
    Luca Maria Aiello\textsuperscript{\rm 1}
}
\affiliations{
    \textsuperscript{\rm 1}IT University of Copenhagen, Denmark\\
    \textsuperscript{\rm 2}University of Bologna, Italy\\
}

\begin{document}

\maketitle

\begin{abstract}
Multimodal Large Language Models (MLLMs) increasingly function as generative search systems that retrieve and synthesize answers from multimedia content, including YouTube videos. Although these systems project authority by citing specific videos as evidence, the extent to which these citations genuinely substantiate the generated claims remains underexplored. We present a large-scale audit of the Gemini 2.5 Pro multimodal search system, analyzing 11,943 claim-video pairs generated across Medical, Economic, and General domains. Through automated verification using three independent LLM judges (87.7\% inter-rater agreement), validated against human annotations, we find that depending on the judge's strictness, between 3.7\% and 18.7\% of video-grounded claims are not supported by their cited sources. The dominant failure modes are not outright contradictions but rather unverifiable specificities and overstated claims, suggesting the system injects precise but ungrounded details from parametric knowledge while citing videos as evidence. Exploratory post-hoc analysis via logistic regression reveals properties associated with these failures: claims departing from source vocabulary ($\beta = -1.6$ to $-3.1$, $p < 0.01$) and claims with low semantic similarity to the video transcript ($\beta = -2.1$ to $-11.6$, $p < 0.01$) are significantly more likely to be unsupported. These findings characterize the current trustworthiness of video-based generative search and highlight the gap between the confidence these systems project and the fidelity of their outputs. The dataset is available at \url{https://anonymous.4open.science/r/icwsm-gemini-audit-04DF}.

\end{abstract}

\section{Introduction}

Multimodal Large Language Models (MLLMs) increasingly function as generative search systems that autonomously retrieve multimedia content and synthesize answers~\citep{ICLR2025_04e65254, zhang2025websearch}.
These systems can process user queries, access the content of online videos, and return factual claims that cite specific videos as supporting evidence.
The impact of this type of multimodal search on online information-seeking outcomes is bound to grow, as users increasingly turn to multimedia-rich social platforms for critical information needs such as medical guidance, financial decisions, and other high-stakes queries~\citep{newman2025digitalnews, pewresearch2025socialmedia}.
A recent audit of over 50,000 health-related queries found that Google's AI Overviews already cite YouTube more than any hospital, government health portal, or medical association~\citep{seranking2026youtube}, making video content from open platforms a core source of sensitive information at scale.

Yet, when these systems generate factual claims and cite specific videos as evidence, the faithfulness of those claims to the source material is not guaranteed, raising fundamental questions about the epistemic reliability of video-based generative search. This is of particular concern given that most users confidently rely on the answers these systems provide~\cite{passi2025addressing,moller2026overreliance}.
Unlike text-based retrieval-augmented generation (RAG), where sources can be readily inspected, video presents unique challenges: transcripts may contain errors originating from Automated Speech Recognition (ASR), visual content may contradict spoken words, and video length makes manual verification impractical.
Verifying a video-grounded claim requires identifying a specific detail within a long video, a cost that arguably very few users would pay for claims that already sound authoritative.
In this sense, the citation functions less as a verifiable reference and more as a trust signal, projecting accountability without necessarily providing it.
While prior work has studied hallucination in text-based LLMs~\citep{ji2023survey} and visual hallucination in videos~\citep{lu2025elvhalluc, gao2025haven}, no study has systematically examined whether video-grounded factual claims produced by generative search systems are actually supported by their cited sources.

To address this gap, we present a large-scale audit of video-grounded claims, in the tradition of black-box algorithmic auditing~\citep{pmlr-v81-buolamwini18a, raji2020audit}.
We use the term \textit{video-grounded} to refer to claims generated from video content via multimodal processing, without presupposing that the cited videos actually support those claims. Evaluating the faithfulness of this grounding is precisely the object of our audit.
We do not assume knowledge of the system's internal retrieval or generation mechanisms; instead, we evaluate its end-to-end functionality.
We analyze 11,943 claim--video pairs across Medical (7,084), Economic (779), and General (4,080) domains. We focus on the Gemini 2.5 Pro model, currently the only MLLM operating as a generative search system with integrated video retrieval capabilities. The Gemini models power Google Search, which dominates the global search engine market~\cite{hu2024market}.

Our audit reveals that a non-trivial fraction of claims lack support from their cited video sources, with rates varying substantially by domain and evaluation stringency.
Crucially, the dominant failure modes are not outright contradictions but rather unverifiable specificities and overstated claims, a pattern consistent with the injection of precise but ungrounded details from parametric knowledge~\citep{xu-etal-2024-knowledge-conflicts}.
These are precisely the failure modes least detectable by users; a claim that adds an exact percentage or a specific entity name not mentioned in the video may in fact sound more authoritative.
Exploratory post-hoc analysis further reveals that claims departing from source vocabulary and claims with low semantic alignment to video transcripts are significantly more likely to be unsupported, suggesting that lexical fidelity and meaning preservation independently contribute to source grounding quality.

\textbf{Contributions:} 
We present the first systematic, large-scale audit of video-grounded claim verifiability in a deployed MLLM generative search system, quantifying the trustworthiness of the user experience across three domains. Through this extensive evaluation, we introduce an error taxonomy with estimates of the frequency of different error categories. Finally, we provide an exploratory regression analysis identifying lexical overlap and transcript semantic similarity as two distinct, complementary properties that characterize when generated claims lack source support. Our contribution is not a comparative claim about video versus text grounding, but a characterization of trustworthiness within a modality where verification is uniquely costly for users.

\section{Related Work}

\textbf{Hallucination in LLMs.} Prior work categorizes hallucinations as intrinsic (contradicting sources) or extrinsic (unverifiable)~\citep{ji2023survey}. Benchmarks like TruthfulQA~\citep{lin2022truthfulqa} evaluate text-only settings, while FActScore~\citep{min2023factscore} introduced atomic-level factuality evaluation by decomposing generations into verifiable claims. Our work examines whether both failure modes manifest when claims are grounded in autonomously retrieved video sources.

\textbf{From RAG to Generative Search.} While RAG faithfulness has been benchmarked in text-based settings~\citep{tamber-etal-2025-benchmarking, jacovi2025facts}, recent audits of deployed generative search engines reveal substantial gaps. Only 51.5\% of generated statements are fully supported by their citations~\citep{liu2023verifiability}, and statement-level audits of deep research agents find citation accuracy ranging from 40--80\% across systems~\citep{venkit2025deeptraceauditingdeepresearch}, with answer engines frequently producing hallucinated or misattributed claims despite appearing authoritative~\citep{10.1145/3715275.3732089}. However, these audits exclusively evaluate text-grounded systems. Video introduces additional complexity through multimodal synthesis and temporal alignment, and no prior work has empirically audited the verifiability of claims produced by video-based generative search systems.

\textbf{Video Hallucination in Multimodal LLMs.} Recent benchmarks have assessed hallucinations in video-based MLLMs across temporal reasoning~\citep{li2025vidhalluc}, semantic aggregation in long videos~\citep{lu2025elvhalluc}, and hallucination causes and formats across multiple models~\citep{gao2025haven}. While these benchmarks measure how well MLLMs \textit{perceive} video content, they do not address whether factual claims \textit{generated and cited by a generative search system} are supported by their cited sources. Our work audits this end-to-end verifiability in a deployed system rather than assessing accuracy in controlled QA settings.

\textbf{Algorithmic Auditing.} Black-box algorithmic auditing evaluates system outputs without access to internal mechanisms~\citep{sandvig2014auditing}. Influential audits have exposed biases in facial recognition~\citep{pmlr-v81-buolamwini18a} and language model toxicity~\citep{gehman2020realtoxicity}. We extend this tradition to the verifiability of video-grounded claims in multimodal generative search.

\section{Methodology}

\subsection{Study Design}
\begin{figure*}[t]
    \centering
    \includegraphics[width=0.85\textwidth]{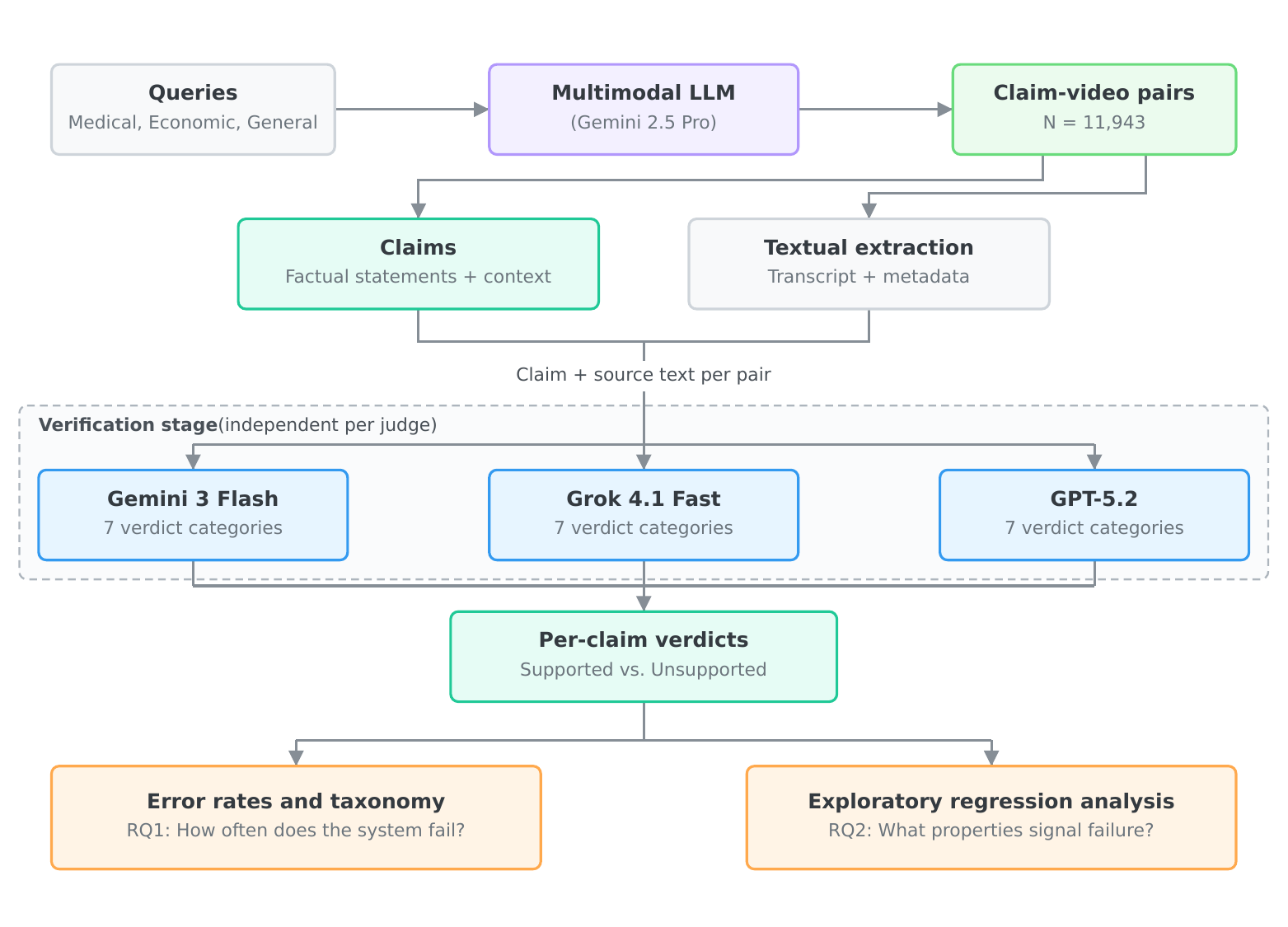}
    \caption{Overview of the auditing pipeline. Queries from three domains are submitted to a multimodal LLM (Gemini 2.5 Pro), which returns claims citing YouTube videos. Each claim--video pair is split into its factual claim and independently extracted textual content: transcript (generated via Whisper ASR), title, description, and upload date. The combined claim and source text are then submitted to three independent LLM judges for verification, yielding per-claim verdicts that inform error-rate analysis and exploratory regression.}
    \label{fig:pipeline}
\end{figure*}

Our audit follows the black-box algorithmic auditing tradition~\citep{pmlr-v81-buolamwini18a}, evaluating only the end-to-end user-facing output of the system without assuming knowledge of its internal retrieval or generation mechanisms. We proceed in three stages. First, we submit queries to Gemini 2.5 Pro, which autonomously retrieves YouTube videos and returns responses containing factual claims linked to specific videos. Second, we extract these claim--video pairs and independently retrieve each video's textual evidence (transcript, title, description, upload date). Third, we submit each claim and its corresponding evidence to a panel of three LLM judges, who independently assess whether the cited video supports the claim. Figure~\ref{fig:pipeline} illustrates this pipeline.

This yields a dataset of 11,943 claim--video pairs 
across three domains: Medical (7,084), Economic (779), and 
General (4,080). In a subsequent exploratory stage, we 
compute structural, lexical, and semantic features of the 
same pairs and use logistic regression to characterize which 
observable properties of these pairs are associated with verification 
failure. The verification and feature analysis stages are 
causally independent: judges have no access to engineered 
features, and features do not inform the judges' decisions.

The following subsections detail each element of this 
pipeline: the query sources and domain selection 
(Data Collection), the multi-judge 
verification protocol, 
human validation of judge reliability, the feature 
engineering process, 
and the statistical analysis framework.

\subsection{Multimodal search model}

We audit Gemini 2.5 Pro, currently the only MLLM operating as a generative search system with integrated video retrieval capabilities. For each query, we prompted the system with a zero-shot instruction requesting claims supported by YouTube videos. The system autonomously searches for and retrieves relevant videos, then generates a response containing factual claims with citations to specific YouTube URLs.

We parsed the returned \texttt{grounding\_metadata} to extract individual factual claims and their corresponding YouTube URLs, capturing the surrounding context for co-reference resolution (e.g., resolving ``it'' to ``the treatment''). Each extracted claim--video pair constitutes one unit of analysis in our audit. We make no assumptions about the system's internal retrieval mechanism, the modalities it processes, or how it selects which videos to cite.

\subsection{Data Collection}
\label{sec:data_collection}

\textbf{Query Sources.} We gathered queries from three domains to evaluate domain-specific failure modes:
\begin{itemize}
    \item \textbf{Medical} (1,000 queries): Health and wellness questions sampled from the Comprehensive Medical Q\&A Dataset~\citep{thedevastator2023medical}.
    \item \textbf{Economic} (109 queries): Financial and market-related questions drawn from the hard web-question subset of FinGPT Search Agents~\citep{10.1145/3677052.3698637}.
    \item \textbf{General} (1,000 queries): Broad knowledge questions sampled from Google's Natural Questions~\citep{kwiatkowski-etal-2019-natural}.
\end{itemize}

\begin{figure*}[t]
    \centering
    \includegraphics[width=\textwidth]{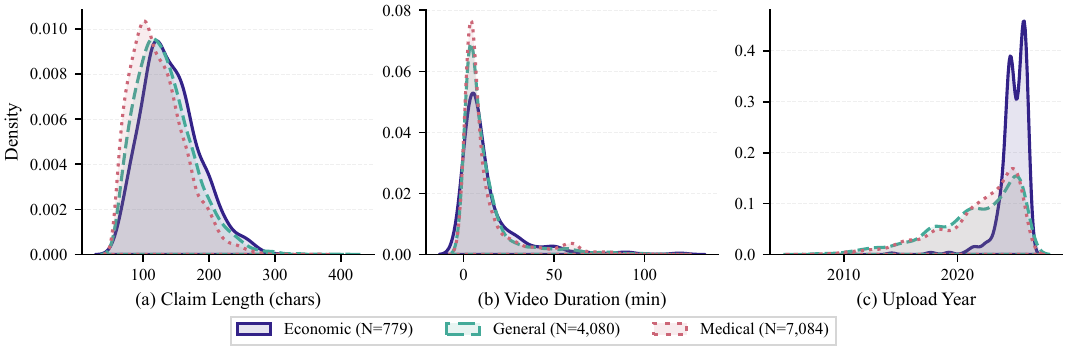}
    \caption{Dataset overview across three domains: (a) claim length distributions; (b) video duration distributions; (c) upload date distributions of cited videos.}
    \label{fig:dataset_overview}
\end{figure*}

\textbf{Filtering.} To help with the process of evaluating standalone facts, we removed claims shorter than 60 characters, retained only claims that cite a single video to isolate source attribution, and excluded failed metadata retrievals, yielding 11,943 distinct claim--video pairs (Medical: 7,084; Economic: 779; General: 4,080). Figure~\ref{fig:dataset_overview} summarizes the resulting dataset across the three domains, showing the distributions of claim length, video duration, and upload dates of cited videos.

\textbf{Textual Extraction.} For each cited video, we independently extracted the textual metadata serving as verification evidence: transcript (generated via the Whisper Automated Speech Recognition model), title, description, video duration, and upload date. We emphasize that these are extracted independently by our auditing pipeline. We make no assumptions about what information the generating system internally accessed or how it processed the cited videos. Our audit does rest on one deliberate assumption: that when the system cites a video as evidence for a claim, a user would reasonably expect that video to support the claim. This reflects the user-facing contract implied by the citation, and it is precisely this contract that we evaluate.

\subsection{Multi-Judge Verification Protocol}
\label{sec:verification}

For each claim--video pair, we assess whether the claim is supported by the cited video's textual evidence. We employ three independent LLM judges: Gemini 3 Flash, Grok 4.1 Fast, and GPT-5.2. Using multiple judges from different families accounts for the inherent ambiguity in verification, where the boundary between a faithful summary and an unsupported extrapolation is often unclear.

Each judge receives a claim, its surrounding context from the generated response (used solely for coreference resolution, e.g., resolving ``it" to ``the treatment"), and the video's textual evidence: transcript, title, description, and upload date. Based on this evidence, the judge classifies the claim into one of seven mutually exclusive categories. A claim is labeled \textsc{supported} if it is a valid summary or extraction of the evidence, else it is labeled as \textsc{unsupported}. Unsupported claims are further specified into sub-categories, including: i) \textsc{contradicted} if it explicitly conflicts with the evidence, ii) \textsc{overstated} if it exaggerates the certainty or scope of what the evidence conveys, iii) \textsc{understated} for claims that significantly downplay the evidence, iv) \textsc{unverifiable} for claims that assert specific details absent from all sources, v) \textsc{temporal\_mismatch} for claims referencing events after the video's upload date, and vi) \textsc{emotion\_distortion} for claims that misinterpret sarcasm or tonal cues.

All judges use the same highly constrained, zero-shot system prompt (temperature = 0), designed to enforce consistent evaluation. The prompt anchors temporal claims to the video's upload date and instructs judges to use the surrounding context only for disambiguation, never as evidence. The full prompt and design variants are discussed in Appendix~\ref{app:verification_prompt}.

\begin{figure}[t]
    \centering
    \includegraphics[width=\columnwidth]{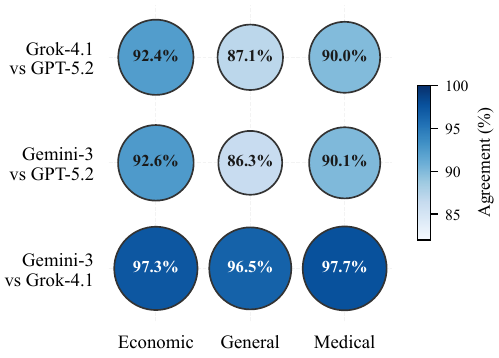}
    \caption{Pairwise agreement rates between LLM judges across domains. Circle size and color encode agreement percentage. Gemini-3 and Grok-4.1 show high concordance ($>$96\%), while pairs involving GPT-5.2 show lower agreement, reflecting its stricter evaluation criteria.}
    \label{fig:judge_agreement}
\end{figure}

\textbf{Inter-Rater Agreement:} The binary agreement across all three judges was Medical 88.9\%, Economic 91.1\%, General 85.0\%, yielding an \textbf{Overall 87.7\%}. Figure~\ref{fig:judge_agreement} shows pairwise agreement rates: Gemini-3 and Grok-4.1 exhibit high concordance ($>$96\%), while pairs involving GPT-5.2 show lower agreement, reflecting its stricter evaluation criteria, a pattern we examine further in the Results.

\subsection{Human Validation}
\label{sec:human_validation}

We validated the LLM judges against human annotations in two rounds. Two annotators first labeled 49 claims together to check that the codebook could be applied consistently; raw agreement was 87.8\%, which we considered sufficient to proceed with single-annotator labeling. Each annotator then independently labeled additional claims, giving us two disjoint samples for the per-judge analysis: Annotator 1 labeled 102 claims (the 49 overlap claims plus 53 additional ones drawn from Gemini-3 Flash's verdicts), and Annotator 2 labeled a separate 101 claims balanced between supported and unsupported verdicts.
Crucially, both samples are balanced in the number of supported and unsupported claims according to Gemini-3 Flash.

Table~\ref{tab:human_validation} shows each judge's agreement with each annotator. Against Annotator 2's balanced sample, all three judges agree with humans at 90.1\%--95.0\%. Gemini-3 Flash missed none of the 46 unsupported claims Annotator 2 identified, and GPT-5.2 missed only one; false alarms were rare (3--5 cases out of 55 supported claims). Agreement is lower against Annotator 1, falling to 80.4\%--90.2\%. GPT-5.2 in particular flagged 12 claims that Annotator 1 considered supported, which we read as the model applying its stricter \textsc{overstated} criterion to borderline cases the human annotator was willing to accept. Grok 4.1 Fast has the weakest agreement against both annotators, and the asymmetry between its false alarms and misses (6 vs.\ 14 against Annotator 1; 3 vs.\ 7 against Annotator 2) is the same leniency pattern visible in its lower main-analysis unsupported-claim rates.

\begin{table}[t]
\centering
\small
\begin{tabular}{lcc}
\toprule
\textbf{Judge} & \textbf{Annotator 1} ($n{=}102$) & \textbf{Annotator 2} ($n{=}101$) \\
\midrule
Gemini 3 Flash & 90.2\% & 95.0\% \\
GPT-5.2        & 82.4\% & 94.1\% \\
Grok 4.1 Fast  & 80.4\% & 90.1\% \\
\bottomrule
\end{tabular}
\caption{LLM judge agreement with human annotations on two disjoint samples, each balanced between supported and unsupported judge verdicts according to Gemini-3 Flash. Annotator 1's sample of 102 claims includes the 49 claims from the inter-annotator reliability subsample.}
\label{tab:human_validation}
\end{table}

\subsection{Feature Extraction for Post-Hoc Analysis}
\label{sec:feature_engineering}

After the verification stage, we conduct an exploratory analysis to understand what observable properties of claims are associated with verification failure. We design features around three hypotheses about how unsupported claims might differ from supported ones.

First, if the generating system fabricates details not present in the source, we would expect unsupported claims to share less vocabulary with the cited video. We capture this through \textit{lexical features} that measure exact overlap between claim terms and source text. Second, even when vocabulary overlaps, a claim may distort the overall meaning of the source. We capture this through \textit{semantic features} that measure embedding-level similarity between the claim and different components of the video's evidence. Third, longer claims contain more surface area for ungrounded details, and longer videos may make faithful grounding harder. We capture this through \textit{structural features} measuring claim length and video duration.

These features are computed independently of the verification stage; the LLM judges have no access to any of them. For lexical features, we employ a deliberately conservative exact-match approach to measure verbatim vocabulary preservation rather than semantic equivalence.

\textbf{Lexical Features.}
For each claim, we parse the claim text using spaCy's \texttt{en\_core\_web\_trf} transformer-based pipeline to extract noun chunks, named entities, and numeric entities. We then measure lexical overlap against the \emph{concatenated} source text (transcript $+$ title $+$ description) via exact substring matching. This concatenation ensures that lexical features capture overlap with any part of the video's textual evidence, whereas semantic features (below) measure similarity to each source independently.

\begin{itemize}
    \item \textit{Noun Overlap}: Proportion of noun chunks in the claim found in the source text. A match is counted if \emph{either} the original surface form or the lemmatized form appears as an exact substring in the source, providing a conservative test that accommodates only morphological variation (e.g., ``treatments'' matching ``treatment'') while rejecting semantic variations.
    \item \textit{Entity Precision}: Proportion of named entities (e.g., persons, organizations, locations) in the claim found in the source via exact substring matching.
    \item \textit{Numeric Precision}: Proportion of numeric entities (\textsc{cardinal}, \textsc{money}, \textsc{percent}, \textsc{quantity}, \textsc{ordinal}) in the claim found in the source via exact substring matching.
\end{itemize}

For all three lexical features, when a claim contains no instances of the relevant category (e.g., no named entities), precision is undefined. We assign a default value of 1.0, treating the absence of verifiable elements as satisfied. This strict criterion measures verbatim vocabulary preservation rather than semantic equivalence, providing a conservative test of extractive behavior. Importantly, only the claim text is parsed with spaCy; source texts are used as raw lowercase strings, ensuring that the matching direction is from claim to source.

\textbf{Semantic Features (BGE-M3 embeddings).}
Unlike the lexical features above, semantic features measure similarity to each source component \emph{independently} using cosine similarity between BGE-M3 normalized embeddings. For longer texts (descriptions and transcripts), we chunk the source into 300-word windows with 50-word overlap and report the maximum similarity between the claim and any chunk.

\begin{itemize}
    \item \textit{Claim-Title Similarity}: Cosine similarity between the claim and the video title.
    \item \textit{Claim-Description Similarity}: Maximum chunk-level similarity with the video description.
    \item \textit{Claim-Transcript Similarity}: Maximum chunk-level similarity with the video transcript.
\end{itemize}
We additionally compute a composite feature, \textit{Maximum Source Similarity} (the maximum of the three individual similarities), for a combined regression model reported in Appendix~\ref{app:combined_regression}.

\textbf{Structural Features:} Claim word count, log-transformed video duration ($\log(1 + \text{duration})$).

\subsection{Regression Analysis}

We model error probability via multivariate logistic regression:
\[ \log\left(\frac{P(\text{error})}{1-P(\text{error})}\right) = \beta_0 + \sum_{i} \beta_i X_i \]

We estimate separate logistic regression models for each domain $\times$ judge combination. All 9 models converged successfully within 6--8 iterations.
We report log-odds coefficients ($\beta$), standard errors, p-values, and pseudo $R^2$.
We compare domain-specific effects by examining coefficient patterns across separate regressions rather than testing cross-domain differences via interaction terms.

We emphasize that the logistic regression serves as an exploratory tool for characterizing which observable claim properties co-occur with verification failure. It is not intended as a predictive classifier or a deployable diagnostic; the pseudo $R^2$ values reflect the expected reality that much of the variance in verification outcomes is driven by claim-specific content that structural features cannot capture. The primary contribution of this audit is the verification itself, quantifying how often and how the system fails users, with the regression providing supplementary insight into the structure of those failures.

\section{Results}

\subsection{Error Rates and Taxonomy}

\begin{figure}[t]
    \centering
    \includegraphics[width=\columnwidth]{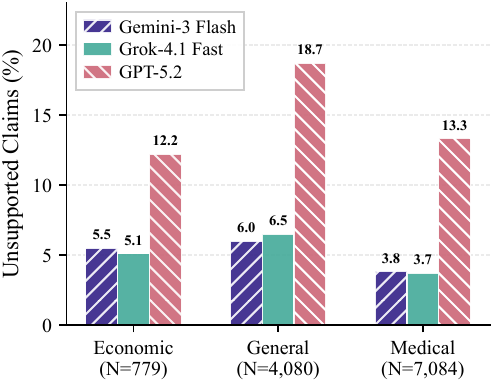}
    \caption{Unsupported Claims (\%) by Domain and Judge. GPT-5.2 shows significantly higher error detection rates, reflecting its stricter threshold for OVERSTATED claims.}
    \label{fig:error_rates}
\end{figure}

Figure~\ref{fig:error_rates} details the baseline error rates by domain and judge. We observe a significant divergence in strictness: GPT-5.2 identifies roughly 2--3$\times$ more unsupported claims (12.2\% -- 18.7\%) than Gemini-3 Flash (3.8\% -- 6.0\%) and Grok-4.1 Fast (3.7\% -- 6.5\%).

This divergence is rooted in how different models categorize verification failures, detailed in Table~\ref{tab:error_distribution}. For GPT-5.2, errors are overwhelmingly classified as \textsc{overstated} (60.3\% to 66.2\% of all detected unsupported claims, across domains), indicating the model frequently flags claims that exaggerate the scope or certainty of the video's content. Conversely, Gemini-3 and Grok-4.1 predominantly flag errors as \textsc{unverifiable} (58.1\% to 82.0\%). This represents a ``specificity hallucination'' pattern, where the generating system injects precise details (e.g., exact dates, unmentioned entity names) that cannot be traced to the cited source. Explicit \textsc{contradicted} errors represent a smaller minority of failures across all judges.

\begin{table*}[t]
\centering
\small
\begin{tabular}{lrrrrrrrrr}
\toprule
 & \multicolumn{3}{c}{\textbf{Economic}} & \multicolumn{3}{c}{\textbf{General}} & \multicolumn{3}{c}{\textbf{Medical}} \\
\cmidrule(lr){2-4} \cmidrule(lr){5-7} \cmidrule(lr){8-10}
\textbf{Label} & Gemini-3 & Grok-4.1 & GPT-5.2 & Gemini-3 & Grok-4.1 & GPT-5.2 & Gemini-3 & Grok-4.1 & GPT-5.2 \\
\midrule
CONTRADICTED & 16.3\% & 15.0\% & 10.5\% & 18.9\% & 13.6\% & 9.0\% & 7.7\% & 7.3\% & 4.1\% \\
OVERSTATED & 7.0\% & 17.5\% & 65.3\% & 6.2\% & 11.4\% & 60.3\% & 8.8\% & 11.1\% & 66.2\% \\
TEMPORAL\_MISMATCH & 14.0\% & 0.0\% & 0.0\% & 3.7\% & 0.0\% & 0.3\% & 0.4\% & 0.0\% & 0.0\% \\
UNDERSTATED & 4.7\% & 0.0\% & 0.0\% & 0.4\% & 0.0\% & 0.1\% & 1.1\% & 0.0\% & 1.5\% \\
UNVERIFIABLE & 58.1\% & 67.5\% & 24.2\% & 70.8\% & 75.0\% & 30.3\% & 82.0\% & 81.7\% & 28.2\% \\
\bottomrule
\end{tabular}
\caption{Error Type Distribution (\%) by Domain and Judge. GPT-5.2 categorizes the majority of errors as \textsc{overstated}, while Gemini and Grok primarily flag \textsc{unverifiable} specificities.}
\label{tab:error_distribution}
\end{table*}

\begin{figure*}[t]
    \centering
    \includegraphics[width=\textwidth]{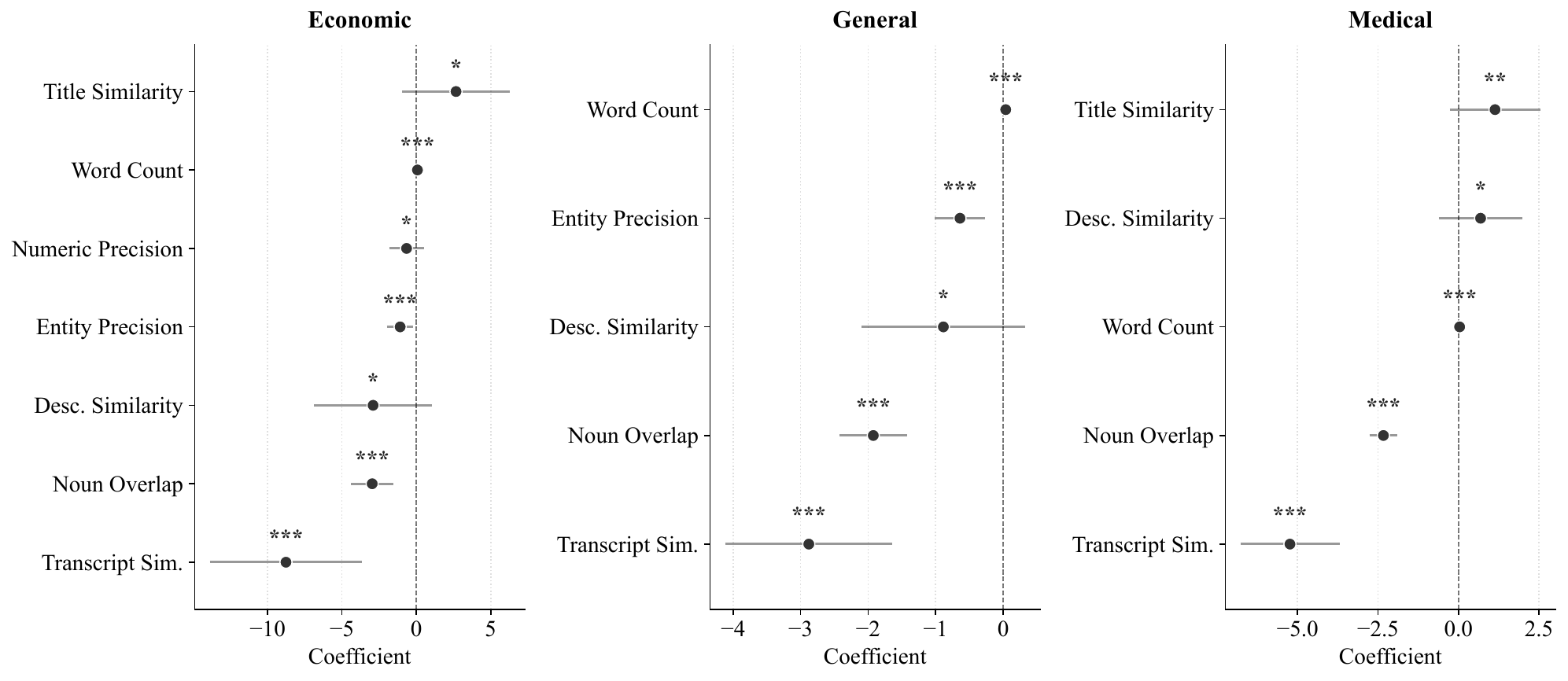}
    \caption{Significant logistic regression coefficients by domain. Each point shows the mean coefficient across three judges; error bars show the averaged 95\% confidence intervals. Only features with at least one significant coefficient ($p < 0.1$) across judges are shown. Significance levels: $^{*}p < 0.1$, $^{**}p < 0.05$, $^{***}p < 0.01$. Transcript similarity and noun overlap are consistently the strongest protective predictors across all domains, while word count is the only consistent risk factor.}
    \label{fig:coefficients}
\end{figure*}

From the user's perspective, these error rates define the trustworthiness of the system. Even under the most lenient judges (Gemini-3 and Grok-4.1), roughly 1 in 20 to 1 in 15 claims citing a YouTube video are not supported by that video. Under the strictest judge (GPT-5.2), this rises to roughly 1 in 5 for the General domain. Moreover, the nature of these failures, predominantly unverifiable specificities and overstated claims rather than obvious contradictions, makes them particularly difficult for users to detect without consulting the original source.

We note that the base error rates for Gemini-3 and Grok-4.1 in the Medical domain are relatively low (3.8\% and 3.7\%, corresponding to 272 and 262 errors out of 7,084 claims).

\subsection{Regression Results}

\begin{figure*}[t]
    \centering
    \includegraphics[width=\textwidth]{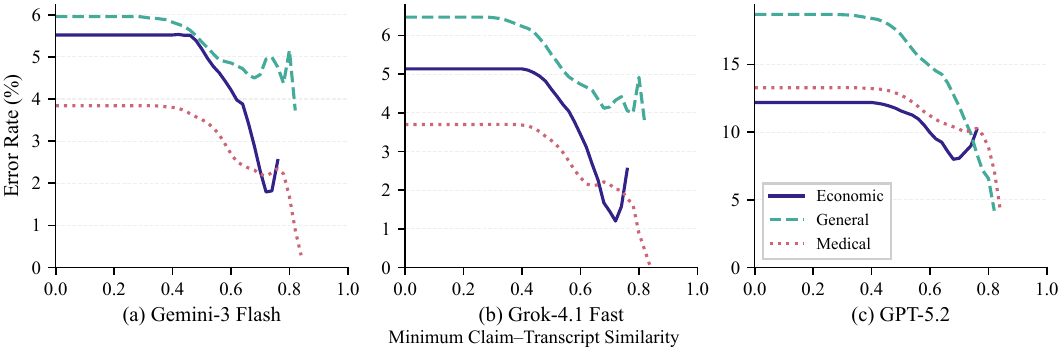}
    \caption{Error rate as a function of claim--transcript 
    similarity. Each point shows the error rate among claims 
    with similarity at or above the given value. The declining trend across all three judges confirms that stronger 
    semantic alignment with transcript content is associated 
    with substantially lower error rates.}
    \label{fig:claim_transcript_sim}
\end{figure*}

Figure~\ref{fig:coefficients} presents the logistic regression coefficients identifying properties associated with verification failure across all three domains. 
Variance Inflation Factors (VIF) are $< 2$ for all features across all models, indicating no problematic multicollinearity.
Two features stand out as consistently strong protective predictors. Noun overlap is significant across all 9 models ($\beta = -1.646$ to $-3.066$, all $p < 0.01$), while transcript similarity emerges as the strongest protective factor overall, with coefficients substantially larger in magnitude than those for any other feature. Figure~\ref{fig:claim_transcript_sim} confirms this, showing error rates declining as transcript similarity increases. This indicates that claims aligned with the spoken content of cited videos are far more likely to be verified as supported; superficial alignment with video metadata alone is insufficient.

\begin{figure}[t]
    \centering
    \includegraphics[width=0.5\textwidth]{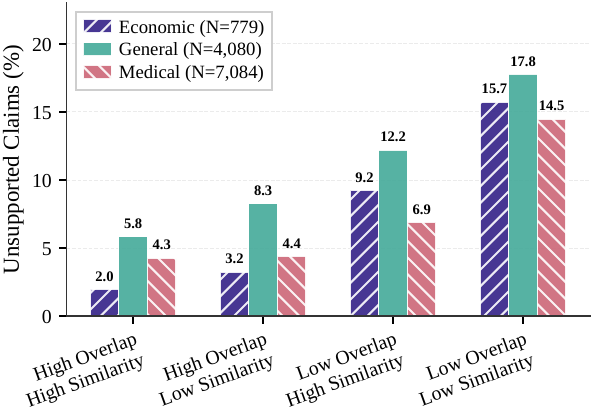}
    \caption{Unsupported claim rates (\%) by noun overlap and transcript similarity. Within each domain, claims are split into High and Low groups by the domain-specific median on each dimension (High: value $\geq$ median; Low: $<$ median). Bars show the unsupported-claim rate within each of the four resulting cells, averaged across the three LLM judges. Error rates increase monotonically as claims depart from the source on either dimension, with the highest rates consistently observed among claims with low values on both.}
    \label{fig:quadrant}
\end{figure}

Although noun overlap and transcript similarity are conceptually related, they capture distinct dimensions of source fidelity: transcript similarity measures semantic alignment via dense embeddings, whereas noun overlap measures verbatim vocabulary preservation via exact matching. A claim can be semantically similar to the transcript while using entirely different words, or reuse key noun phrases while distorting overall meaning. The low multicollinearity between these features (VIF $< 2$ across all models) confirms their independence, and both are independently significant in the regression. Figure~\ref{fig:quadrant} shows this complementarity descriptively: the highest error rates concentrate among claims that are distant from the source on both dimensions, while claims close on either dimension show substantially lower rates. This pattern holds across all three domains.

Title similarity, by contrast, shows weak and inconsistent effects, with occasionally positive (risk-increasing) coefficients, suggesting that the primary correlate of verification failure is insufficient alignment with transcript content rather than over-reliance on titles. Claim word count consistently predicts verification failure, with longer claims providing more surface area for ungrounded details. Entity precision shows significant protective effects in the Economic and General domains but not in Medical, where 79.2\% of claims contain no named entities, leaving too sparse data for the regression to detect an effect.

\section{Discussion}

\subsection{Trustworthiness of Multimodal AI Search}

Our audit shows that the system produces claims supported by the cited video in the large majority of cases. Yet, a non-trivial fraction (from 1 in 25 to 1 in 5 depending on domain and evaluation stringency), are not supported by the video they cite. From the user's perspective, these unsupported claims are indistinguishable from supported ones: they are presented with the same confidence, the same citation format, and the same implicit guarantee that the source substantiates the claim.

The nature of these failures is particularly critical, given the trust that users have in the system's responses. The system predominantly produces claims containing unverifiable specificities and overstated conclusions rather than outright contradictions. These are failure modes that are hardly detectable by users: a claim that adds an exact percentage, a specific date, or a named entity not mentioned in the video \textit{sounds} more authoritative, not less.

This trust problem is compounded by the verification asymmetry inherent to video citations. When a generative search system cites a webpage, a user can verify the claim in seconds by searching the page for the relevant passage. When the system cites a YouTube video, the same verification requires scrubbing through content that may be minutes or hours long, with no equivalent of a text search for spoken content. This means that even a modest error rate in video-grounded search is functionally more harmful than a comparable rate in text-grounded search, because the practical cost of catching errors is orders of magnitude higher. The error rates we report should be interpreted in this context: the relevant question is not whether they are higher or lower than text-based benchmarks, but whether they are acceptable given that users have essentially no realistic opportunity to verify them.

\subsection{Characterizing Where Failures Concentrate}

Our analysis reveals a consistent pattern across all nine domains--judge models: claims whose observable properties indicate closeness to the source, both lexically and semantically, are substantially more likely to be supported. Noun overlap and transcript similarity emerge as the two strongest associated features, operating independently along complementary dimensions: vocabulary preservation and meaning alignment. Title similarity shows weak and inconsistent effects instead, and claim verbosity consistently increases error risk. 

These associations are consistent with a specific failure mechanism: when the system departs from what the video actually says, whether by paraphrasing too liberally, injecting details from parametric knowledge, or overstating the source's claims, the resulting output is more likely to be unsupported. We emphasize that this characterization is correlational; we cannot determine whether lexical departure \textit{causes} verification failure or merely co-occurs with it. Nonetheless, the consistency across all nine models, including GPT-5.2 which applies the strictest evaluation criteria, suggests a robust empirical regularity.

\subsection{Social Implications}

Beyond algorithmic benchmarking, our findings carry significance for computational social science more broadly. As MLLMs are deployed as generative search systems that autonomously retrieve and synthesize from multimedia sources, their propensity to inject unverifiable specificities poses a novel vector for misinformation. The generative search paradigm is particularly concerning because it compounds two trust signals: the authority of a search engine with the fluency of a language model. Users who trust an AI search agent's summary of a lengthy video are highly unlikely to manually verify the video transcript, underscoring the importance of systematic auditing such as the one presented here.

\subsection{Visual Grounding as Complementary Evidence}

A limitation of our main analysis is that verification relies on textual evidence (transcripts, titles, descriptions). Some claims may be supported by information present only in visual form (e.g., on-screen graphics, demonstrations) that is not captured in transcripts, and such claims would be classified as \textsc{unverifiable} under our text-only protocol. To assess this limitation, we conducted a supplementary analysis on a stratified subsample of 1,028 claim--video pairs, balanced between supported and unsupported claims, providing Gemini-3 Flash with direct access to video content via video tokens alongside transcripts and metadata.

Of the 997 pairs remaining after API errors, 624 (62.6\%) were classified as supported under the video-token condition, compared to the designed 50\% baseline. This indicates that approximately 125 claims shifted from unsupported to supported when visual evidence was available, while the majority of unsupported claims remained unsupported even with full video access. Visual modalities were cited in 16.0\% of decisions, suggesting that these modalities provide supplementary verification evidence in a subset of cases. Because this subsample was stratified rather than randomly drawn, these rates should not be interpreted as population-level estimates.

\subsection{Judge Disagreement}

GPT-5.2 identified substantially more errors than Gemini-3 Flash and Grok-4.1 Fast, driven by stricter \textsc{overstated} criteria. This reflects some level of ambiguity in verification: a claim may be directionally correct but exaggerate the certainty or scope of the source. Our multi-judge approach surfaces rather than hides this ambiguity, and the consistency of our key findings across all three judges strengthens our conclusions.

\subsection{Limitations}

Our audit has several limitations that should be considered when interpreting the results. We audit the end-to-end output of Gemini 2.5 Pro without knowledge of its internal retrieval or generation mechanisms.
This means we cannot determine \textit{why} certain claims depart from source content, nor disentangle retrieval failures (citing an irrelevant video) from generation failures (producing an unfaithful claim from a relevant video). Furthermore, because Gemini 2.5 Pro is currently the only MLLM operating as a generative search system with integrated video retrieval, our findings characterize this specific system's behavior; other systems may exhibit different patterns as they gain similar capabilities.

Our main analysis verifies claims against textual metadata only, and as our supplementary analysis shows, visual modalities provide additional evidence in approximately 16\% of cases. Our lexical features use exact substring matching, which does not capture semantic equivalence (e.g., ``heart attack" vs.\ ``cardiac arrest"), potentially underestimating the protective association of semantic fidelity.

While our three judges show high binary agreement, they may share systematic biases. One judge (Gemini 3 Flash) belongs to the same model family as the generating system; however, the consistency across all three judges, including GPT-5.2 with the strictest criteria, mitigates this concern. Human validation on 49 flagged claims confirmed strong alignment with human consensus for Gemini 3 Flash and GPT-5.2, with false alarms limited to two or fewer cases per judge. Finally, our regression identifies associations, not causal mechanisms, and the modest pseudo $R^2$ values confirm that the engineered features capture only a portion of the variance in verification outcomes.

\section{Conclusion}

We present the first large-scale forensic audit of video-grounded claims produced by an MLLM generative search system, analyzing 11,943 claims across medical, economic, and general domains. Treating the system as a black box and evaluating the end-to-end user experience, our audit reveals that between 3.7\% and 18.7\% of claims are not supported by their cited video sources. The dominant failure modes are not outright contradictions but unverifiable specificities and overstated claims, failures that are particularly insidious because they are difficult for users to detect without consulting the original source.

Post-hoc analysis identifies two complementary properties associated with verification failure: claims that depart from source vocabulary and claims with low semantic similarity to video transcripts are both significantly more likely to be unsupported. These two factors operate independently, suggesting that vocabulary fidelity and semantic alignment capture complementary dimensions of source grounding quality. Supplementary analysis with visual grounding confirms that the dominant verification dynamics remain transcript-driven, while recovering a subset of claims supported by on-screen content not captured in transcripts.

As MLLMs increasingly mediate access to video content through autonomous search and retrieval, the trust relationship between users and these systems becomes uniquely opaque: unlike text citations that can be verified in seconds, video citations impose a verification cost that virtually no user will pay. Systematic auditing of the kind presented here is essential for maintaining epistemic integrity in this emerging modality. Our findings do not claim that video grounding is more or less faithful than text grounding, but rather that video-grounded search creates a distinct trust problem that warrants dedicated scrutiny.

\section{Data Availability}
\label{sec:data_availability}

To support reproducibility, we release the audit data at \url{https://anonymous.4open.science/r/icwsm-gemini-audit-04DF}, organized into four files.

The main dataset is split by domain into three files, together covering all 11,943 claim--video pairs. Each row provides the original query and its domain, the extracted claim and its surrounding response text, the cited video's metadata (url, title, description, upload date) and Whisper transcript, and, for each of the three judges, both the binary verdict the binary verdict (\emph{YES} for supported, \emph{NO} for unsupported) and the full output with the assigned category, primary source, and reasoning. The same files also contain the engineered features used in the regression analysis: lexical overlap (noun\_overlap, entity\_precision, numeric\_precision), BGE-M3 cosine similarities between the claim and each video source (sim\_claim\_title, sim\_claim\_description, sim\_claim\_transcript, sim\_claim\_source\_max), and structural features (claim\_word\_count, log\_duration).

A separate file releases the 203 human-annotated claims, with the same schema as the main files plus the two annotators' labels, confidence ratings, and notes. The 49 claims labeled by both annotators carry values for both; the remaining claims are single-annotated.

In compliance with YouTube's Terms of Service, we redistribute only video identifiers and metadata; original video files are not re-hosted.

\bibliography{aaai25}

\subsection{Paper Checklist}

\begin{enumerate}

\item For most authors.
\begin{enumerate}
    \item  Would answering this research question advance science without violating social contracts, such as violating privacy norms, perpetuating unfair profiling, exacerbating the socio-economic divide, or implying disrespect to societies or cultures?
    \answerYes{Yes, the study uses publicly available video metadata and AI-generated outputs; no private data, demographic profiling, or individually identifiable users are involved.}
  \item Do your main claims in the abstract and introduction accurately reflect the paper's contributions and scope?
    \answerYes{Yes.}
   \item Do you clarify how the proposed methodological approach is appropriate for the claims made? 
    \answerYes{Yes.}
   \item Do you clarify what are possible artifacts in the data used, given population-specific distributions?
    \answerYes{Yes.}
  \item Did you describe the limitations of your work?
    \answerYes{Yes.}
  \item Did you discuss any potential negative societal impacts of your work?
    \answerYes{Yes, we discussed it in the social impacts subsection.}
      \item Did you discuss any potential misuse of your work?
    \answerNA{NA}
    \item Did you describe steps taken to prevent or mitigate potential negative outcomes of the research, such as data and model documentation, data anonymization, responsible release, access control, and the reproducibility of findings?
    \answerYes{Yes.}
  \item Have you read the ethics review guidelines and ensured that your paper conforms to them?
    \answerYes{Yes.}
\end{enumerate}

\item Additionally, if your study involves hypotheses testing.
\begin{enumerate}
  \item Did you clearly state the assumptions underlying all theoretical results?
    \answerNA{Our study is an exploratory audit, not a
hypothesis-testing study.}
  \item Have you provided justifications for all theoretical results?
    \answerNA{NA}
  \item Did you discuss competing hypotheses or theories that might challenge or complement your theoretical results?
    \answerNA{NA}
  \item Have you considered alternative mechanisms or explanations that might account for the same outcomes observed in your study?
    \answerNA{NA}
  \item Did you address potential biases or limitations in your theoretical framework?
    \answerNA{NA}
  \item Have you related your theoretical results to the existing literature in social science?
    \answerNA{NA}
  \item Did you discuss the implications of your theoretical results for policy, practice, or further research in the social science domain?
    \answerNA{NA}
\end{enumerate}

\item Additionally, if you are including theoretical proofs.
\begin{enumerate}
  \item Did you state the full set of assumptions of all theoretical results?
    \answerNA{NA}
	\item Did you include complete proofs of all theoretical results?
    \answerNA{NA}
\end{enumerate}

\item Additionally, if you ran machine learning experiments.
\begin{enumerate}
  \item Did you include the code, data, and instructions needed to reproduce the main experimental results (either in the supplemental material or as a URL)?
    \answerYes{Yes. The Data Availability section provides the dataset,
engineered features, and judge outputs. The full verification
prompt is reproduced in the Appendix.}
  \item Did you specify all the training details (e.g., data splits, hyperparameters, how they were chosen)?
    \answerNo{No, because we didn't conduct any training.}
     \item Did you report error bars (e.g., with respect to the random seed after running experiments multiple times)?
    \answerNA{NA}
	\item Did you include the total amount of compute and the type of resources used (e.g., type of GPUs, internal cluster, or cloud provider)?
    \answerNA{NA}
     \item Do you justify how the proposed evaluation is sufficient and appropriate to the claims made? 
    \answerYes{Yes.}
     \item Do you discuss what is ``the cost`` of misclassification and fault (in)tolerance?
    \answerYes{Yes.}
  
\end{enumerate}

\item Additionally, if you are using existing assets (e.g., code, data, models) or curating/releasing new assets, without compromising anonymity.
\begin{enumerate}
  \item If your work uses existing assets, did you cite the creators?
    \answerYes{Yes.}
  \item Did you mention the license of the assets?
    \answerNA{NA}
  \item Did you include any new assets in the supplemental material or as a URL?
    \answerYes{Yes}
  \item Did you discuss whether and how consent was obtained from people whose data you're using/curating?
    \answerNA{NA}
  \item Did you discuss whether the data you are using/curating contains personally identifiable information or offensive content?
    \answerNA{NA}
\item If you are curating or releasing new datasets, did you discuss how you intend to make your datasets FAIR (see \citet{fair})?
\answerYes{Yes.}
\item If you are curating or releasing new datasets, did you create a Datasheet for the Dataset? 
\answerNo{Not yet, but we will do it before the dataset is publicly released.}
\end{enumerate}

\item Additionally, if you used crowdsourcing or conducted research with human subjects, without compromising anonymity.
\begin{enumerate}
  \item Did you include the full text of instructions given to participants and screenshots?
    \answerYes{Yes.}
  \item Did you describe any potential participant risks, with mentions of Institutional Review Board (IRB) approvals?
    \answerNA{NA}
  \item Did you include the estimated hourly wage paid to participants and the total amount spent on participant compensation?
    \answerNA{NA}
   \item Did you discuss how data is stored, shared, and deidentified?
   \answerYes{Yes}
\end{enumerate}

\end{enumerate}

\appendix

\section{Appendix}

\subsection{Verification Prompt}
\label{app:verification_prompt}

The following prompt was used for all three LLM judges. It enforces a strict evidence hierarchy prioritizing the transcript, followed by the description, and finally the title. We also tested a variant without this hierarchy and found consistent results, supporting the robustness of the protocol.

\begin{lstlisting}
You are an expert scientific fact-checker. Your task is to verify whether a specific TARGET_CLAIM is textually, proportionally, and temporally supported by a YouTube video's metadata.

You must act as a "Reasonable Textual Auditor." You cannot see the visuals, but you should use logic to infer the context from the text.

### 1. CONTEXTUAL DISAMBIGUATION PROTOCOL (Crucial)
You are provided with `GENERATED_CONTEXT_FOR_DISAMBIGUATION` (the full text response) and a specific `TARGET_CLAIM` (SPECIFIC part from the full text response).
* The Problem: The `TARGET_CLAIM` may contain pronouns ("This," "It," "The method") whose definitions are in the surrounding text.
* The Rule: Use the `GENERATED_CONTEXT` ONLY to identify what the subject refers to.
    * Example: Context says "Use Python for data." Claim says "It is slow." -> You must interpret Claim as "Python is slow."
* The Constraint: The `GENERATED_CONTEXT` is NOT evidence. Never use it to support the claim. You must verify the interpreted claim ONLY against the provided VIDEO_METADATA (Transcript, Description, Title, Upload Date).

### 2. SOURCE ANALYSIS RULES
1.  HIERARCHY: TRANSCRIPT > DESCRIPTION > TITLE.
    * If Title contradicts Transcript (clickbait), trust Transcript.
    * If Transcript is silent/ambiguous, you MAY use Title/Description to confirm the general topic.
2.  DATE ANCHORING: Use UPLOAD_DATE as the absolute anchor.
3.  DE-NOISING: The claim may contain conversational artifacts like "In another video," "He mentions," or "Conversely." IGNORE these phrases. Focus on the atomic fact.

### 3. LOGIC GATES (Judgement Rules)
* The "Pronoun" Check: After defining the subject using the Context (Protocol 1), check if the METADATA attributes the claimed property to that subject.
    * Match: Metadata supports the fact (explicitly or via logical implication) -> SUPPORTED.
    * Conflict: Metadata contradicts the fact -> CONTRADICTED.
    * Silence: Metadata neither mentions nor implies the subject/fact -> UNVERIFIABLE.
* The "Gist" Rule: If the claim accurately summarizes the specific topic, action, or prediction of the video, mark SUPPORTED.
* The "Inference" Limit: Logical consequences (Walking + Woods = Hiking) are SUPPORTED. Specific missing data (Colors, Numbers) are UNVERIFIABLE.
* The "Retrospective" Exception: Past-tense descriptions of predictions made in the video (e.g., "It narrowed to April") are SUPPORTED if the prediction exists in the text.

### 4. CLASSIFICATION LABELS (Select Exactly One)

SUPPORTED
The claim (as resolved by context) is a valid summary, inference, or extraction of the video evidence.

OVERSTATED
The claim is rooted in the video text but exaggerates certainty or scope.
* Criteria: Changing "some" to "all," removing caveats (e.g., changing "might" to "is"), or stating a hypothesis as a proven fact.

UNDERSTATED
The claim significantly downplays or minimizes the impact, urgency, or certainty found in the video text.

CONTRADICTED
The claim explicitly conflicts with the video text.
* Criteria: The text states X is False, but the Claim states X is True (or vice versa).

UNVERIFIABLE
The claim asserts SPECIFIC missing details (Numbers, Colors, Proper Nouns) absent from all video sources.
* Note: If the context defines "This" as "iPhone 15", but the Metadata only says "Phone" (no model number), this is UNVERIFIABLE.

EMOTION_DISTORTION
The claim misinterprets the source meaning due to a failure to detect sarcasm, irony, or tonal negation in the transcript.

TEMPORAL_MISMATCH
The claim places the video content in a time period that did not exist when the video was uploaded.
* Criteria: Discussing events that happened after the UPLOAD_DATE (e.g., a 2021 video discussing 2024 election results).

### 5. ATTRIBUTION LOGIC (How to cite Primary Source)
You must cite the Single Most Determinant Source that drove your decision:
* If SUPPORTED/CONTRADICTED/OVERSTATED: Cite the source containing the evidence (usually TRANSCRIPT).
* If TEMPORAL_MISMATCH: You MUST cite UPLOAD_DATE.
* If UNVERIFIABLE: You MUST cite NONE (since the info is absent).
* If relying on Title vs Transcript conflict: Cite TRANSCRIPT.
* If relying on Description for a name/entity: Cite DESCRIPTION.
* If relying on Title because Transcript is silent: Cite TITLE.

### 6. OUTPUT FORMAT
Provide the output as four distinct lines of text. Do not use JSON.

Answer: YES / NO - [LABEL]
Primary Source: [TRANSCRIPT / DESCRIPTION / TITLE / UPLOAD_DATE / NONE]
Reason: [One concise sentence explaining the logical derivation.]
Detail: [Comparison snippet: "Source: quote (or date)" vs "Claim: quote (or date)"]
\end{lstlisting}

\FloatBarrier
\subsection{Regression Tables}
\label{app:combined_regression}

This appendix reports the full regression results underlying Figure~\ref{fig:coefficients} in the main text. Table~\ref{tab:combined_regression} presents coefficients for all individual similarity measures (title, description, transcript) alongside lexical and structural features. Table~\ref{tab:combined_regression_robustness} replicates this analysis as a robustness check, replacing the three individual similarity measures with a single composite feature: the maximum cosine similarity across all three sources.

\begin{table*}[t]
\centering
\caption{Logistic Regression Results: Individual Similarity Measures Across Domains. Features are computed independently of and after the verification stage; judges have no access to these features.}
\label{tab:combined_regression}
\resizebox{\textwidth}{!}{%
\begin{tabular}{@{}l*{3}{r}*{3}{r}*{3}{r}@{}}
\toprule
 & \multicolumn{3}{c}{\textbf{Economic (N=779)}} & \multicolumn{3}{c}{\textbf{General (N=4,080)}} & \multicolumn{3}{c}{\textbf{Medical (N=7,084)}} \\
\cmidrule(lr){2-4} \cmidrule(lr){5-7} \cmidrule(lr){8-10}
 & Gemini-3 & Grok-4.1 & GPT-5.2 & Gemini-3 & Grok-4.1 & GPT-5.2 & Gemini-3 & Grok-4.1 & GPT-5.2 \\
\midrule
Entity Precision & $-$1.076$^{**}$ & $-$1.088$^{**}$ & $-$1.101$^{***}$ & $-$0.765$^{***}$ & $-$0.682$^{***}$ & $-$0.469$^{***}$ & 0.021 & $-$0.218 & 0.135 \\
 & (0.484) & (0.498) & (0.377) & (0.214) & (0.209) & (0.150) & (0.403) & (0.381) & (0.249) \\[3pt]
Numeric Precision & $-$0.874 & $-$1.150$^{*}$ & 0.055 & 0.261 & $-$0.199 & $-$0.089 & $-$0.690 & $-$0.190 & $-$0.284 \\
 & (0.607) & (0.620) & (0.544) & (0.302) & (0.262) & (0.200) & (0.471) & (0.480) & (0.312) \\[3pt]
Noun Overlap & $-$3.066$^{***}$ & $-$2.879$^{***}$ & $-$2.961$^{***}$ & $-$1.928$^{***}$ & $-$2.196$^{***}$ & $-$1.646$^{***}$ & $-$2.590$^{***}$ & $-$2.659$^{***}$ & $-$1.750$^{***}$ \\
 & (0.800) & (0.812) & (0.558) & (0.296) & (0.287) & (0.179) & (0.260) & (0.266) & (0.143) \\[3pt]
Title Similarity & 3.244 & 2.067 & 2.690$^{*}$ & 0.040 & 0.031 & $-$0.194 & 1.060 & 1.810$^{**}$ & 0.538 \\
 & (1.974) & (2.119) & (1.428) & (0.736) & (0.707) & (0.459) & (0.835) & (0.852) & (0.473) \\[3pt]
Desc. Similarity & $-$3.275 & $-$2.683 & $-$2.764$^{*}$ & $-$1.060 & $-$0.731 & $-$0.866$^{*}$ & 0.761 & 0.458 & 0.843$^{*}$ \\
 & (2.174) & (2.371) & (1.539) & (0.710) & (0.699) & (0.444) & (0.768) & (0.787) & (0.431) \\[3pt]
Transcript Sim. & $-$8.080$^{***}$ & $-$11.555$^{***}$ & $-$6.655$^{***}$ & $-$2.108$^{***}$ & $-$3.040$^{***}$ & $-$3.487$^{***}$ & $-$5.461$^{***}$ & $-$6.638$^{***}$ & $-$3.612$^{***}$ \\
 & (2.833) & (3.013) & (2.000) & (0.741) & (0.691) & (0.466) & (0.903) & (0.917) & (0.535) \\[3pt]
Word Count & 0.073$^{***}$ & 0.066$^{***}$ & 0.082$^{***}$ & 0.040$^{***}$ & 0.028$^{***}$ & 0.047$^{***}$ & 0.040$^{***}$ & 0.035$^{***}$ & 0.031$^{***}$ \\
 & (0.024) & (0.025) & (0.018) & (0.009) & (0.009) & (0.006) & (0.010) & (0.010) & (0.006) \\[3pt]
Log(Duration) & $-$0.058 & 0.167 & 0.029 & 0.024 & $-$0.079 & $-$0.037 & $-$0.101 & $-$0.102 & $-$0.049 \\
 & (0.172) & (0.176) & (0.119) & (0.066) & (0.065) & (0.042) & (0.067) & (0.068) & (0.037) \\
\midrule
Pseudo $R^2$ & 0.177 & 0.203 & 0.162 & 0.068 & 0.089 & 0.079 & 0.091 & 0.103 & 0.050 \\
\bottomrule
\multicolumn{10}{l}{\footnotesize Standard errors in parentheses. $^{*}p<0.1$; $^{**}p<0.05$; $^{***}p<0.01$} \\
\end{tabular}%
}
\end{table*}

\begin{table*}[!htbp]
\centering
\caption{Logistic Regression Results: Combined Maximum Source Similarity (Robustness Check). This table replicates the analysis in Table~\ref{tab:combined_regression} but replaces the three individual similarity measures (title, description, transcript) with a single composite feature: the maximum cosine similarity across all three sources.}
\label{tab:combined_regression_robustness}
\resizebox{\textwidth}{!}{%
\begin{tabular}{@{}l*{3}{r}*{3}{r}*{3}{r}@{}}
\toprule
 & \multicolumn{3}{c}{\textbf{Economic (N=779)}} & \multicolumn{3}{c}{\textbf{General (N=4,080)}} & \multicolumn{3}{c}{\textbf{Medical (N=7,084)}} \\
\cmidrule(lr){2-4} \cmidrule(lr){5-7} \cmidrule(lr){8-10}
 & Gemini-3 & Grok-4.1 & GPT-5.2 & Gemini-3 & Grok-4.1 & GPT-5.2 & Gemini-3 & Grok-4.1 & GPT-5.2 \\
\midrule
Entity Precision & $-$1.068$^{**}$ & $-$1.094$^{**}$ & $-$1.131$^{***}$ & $-$0.753$^{***}$ & $-$0.673$^{***}$ & $-$0.484$^{***}$ & $-$0.045 & $-$0.302 & 0.107 \\
 & (0.481) & (0.497) & (0.374) & (0.213) & (0.208) & (0.150) & (0.399) & (0.377) & (0.248) \\[3pt]
Numeric Precision & $-$0.880 & $-$1.106$^{*}$ & 0.066 & 0.208 & $-$0.276 & $-$0.157 & $-$0.614 & $-$0.106 & $-$0.242 \\
 & (0.592) & (0.607) & (0.532) & (0.304) & (0.265) & (0.200) & (0.467) & (0.477) & (0.310) \\[3pt]
Noun Overlap & $-$3.165$^{***}$ & $-$2.933$^{***}$ & $-$3.053$^{***}$ & $-$1.898$^{***}$ & $-$2.170$^{***}$ & $-$1.683$^{***}$ & $-$2.551$^{***}$ & $-$2.579$^{***}$ & $-$1.717$^{***}$ \\
 & (0.803) & (0.819) & (0.558) & (0.291) & (0.282) & (0.176) & (0.259) & (0.263) & (0.142) \\[3pt]
Max Source Sim. & $-$7.292$^{***}$ & $-$11.617$^{***}$ & $-$5.506$^{***}$ & $-$3.973$^{***}$ & $-$4.858$^{***}$ & $-$4.883$^{***}$ & $-$4.382$^{***}$ & $-$5.770$^{***}$ & $-$2.854$^{***}$ \\
 & (2.433) & (2.649) & (1.750) & (0.896) & (0.867) & (0.563) & (0.841) & (0.871) & (0.473) \\[3pt]
Word Count & 0.067$^{***}$ & 0.063$^{**}$ & 0.074$^{***}$ & 0.045$^{***}$ & 0.034$^{***}$ & 0.051$^{***}$ & 0.041$^{***}$ & 0.038$^{***}$ & 0.032$^{***}$ \\
 & (0.024) & (0.025) & (0.018) & (0.009) & (0.010) & (0.006) & (0.010) & (0.011) & (0.006) \\[3pt]
Log(Duration) & $-$0.101 & 0.142 & $-$0.009 & 0.009 & $-$0.117$^{*}$ & $-$0.069$^{*}$ & $-$0.164$^{**}$ & $-$0.173$^{***}$ & $-$0.094$^{***}$ \\
 & (0.172) & (0.177) & (0.118) & (0.066) & (0.065) & (0.042) & (0.064) & (0.066) & (0.035) \\
\midrule
Pseudo $R^2$ & 0.169 & 0.204 & 0.151 & 0.072 & 0.094 & 0.080 & 0.086 & 0.099 & 0.048 \\
\bottomrule
\multicolumn{10}{l}{\footnotesize Standard errors in parentheses. $^{*}p<0.1$; $^{**}p<0.05$; $^{***}p<0.01$} \\
\end{tabular}%
}
\end{table*}

\subsection{Examples}
\label{app:representative_examples}
 
\noindent The following examples provide the full claim text, relevant transcript excerpt, and judge reasoning for representative failure cases.

\vspace{6pt}
\noindent\textbf{Example 1 --- Imprecise statistics}
\begin{itemize}
\setlength{\itemsep}{1pt}
\setlength{\parskip}{0pt}
\item \textbf{Video:} Retinoblastoma: Everything You Need To Know (uploaded 2023-01-26)
\item \textbf{Claim:} \textit{``About 40\% of retinoblastoma cases are due to a mutation in the RB1 gene that is passed down from a parent.''}
\item \textbf{Transcript excerpt:} ``In approximately 40\% of the cases, retinoblastoma is a result of a faulty gene that affects the eyes bilaterally. It can be both inherited and acquired. In the other 60\% of cases that are not due to a faulty gene, only one eye is affected.''
\item \textbf{Verdicts:} Gemini-3: \textsc{overstated}; GPT5: \textsc{overstated}; Grok: \textsc{unverifiable}
\item \textbf{Reasoning (Gemini-3 Flash):}
The transcript states that the 40\% of cases involving a faulty gene can be ``both inherited and acquired,'' whereas the claim attributes the entire 40\% solely to being ``passed down from a parent.
\end{itemize}

\vspace{6pt}
\noindent\textbf{Example 2 --- Misattribution}
\begin{itemize}
\setlength{\itemsep}{1pt}
\setlength{\parskip}{0pt}
\item \textbf{Video:} The Hidden Nations of Asia That Few People Know About (uploaded 2018-06-01)
\item \textbf{Claim:} \textit{``In Central Asia, Turkmenistan's governance drew authoritarian accusations from the West''}
\item \textbf{Transcript excerpt:} ``[Tajikistan] only recently has it achieved a little political stability, which comes with a lot of authoritarian accusations by the West''
\item \textbf{Verdicts:} Gemini-3: \textsc{contradicted}; GPT5: \textsc{unverifiable}; Grok: \textsc{contradicted}
\item \textbf{Reasoning (Gemini-3 Flash):}
The transcript explicitly attributes the phrase ``authoritarian accusations by the West'' to Tajikistan, whereas the claim incorrectly attributes it to Turkmenistan.
\end{itemize}

\vspace{6pt}
\noindent\textbf{Example 3 --- Event sequence error}
\begin{itemize}
\setlength{\itemsep}{1pt}
\setlength{\parskip}{0pt}
\item \textbf{Video:} Sabeer Bhatia @Hotmail.com: The Man Behind World's First Web Based Email (uploaded 2022-08-22)
\item \textbf{Claim:} \textit{``Prior to Firepower Systems, Jack Smith had also been a colleague of Bhatia's at Apple.''}
\item \textbf{Transcript excerpt:} ``After spending almost a year at Apple, he joined a startup called the Firepower Systems, where he met his Hotmail co-founder, Jack Smith''
\item \textbf{Verdicts:} Gemini-3: \textsc{contradicted}; GPT5: \textsc{contradicted}; Grok: \textsc{contradicted}
\item \textbf{Reasoning (Gemini-3 Flash):} The transcript explicitly states that Bhatia met Jack Smith at Firepower Systems, which occurred after Bhatia's tenure at Apple.
\end{itemize}

\vspace{6pt}
\noindent\textbf{Example 4 --- Made up number}
\begin{itemize}
\setlength{\itemsep}{1pt}
\setlength{\parskip}{0pt}
\item \textbf{Video:} New Autism Research Reveals 4 Genetic Types - Which One Are You? (uploaded 2025-08-10)
\item \textbf{Claim:} \textit{``A subtype characterized by significant developmental delays and challenges with motor skills and daily living (about 14\%)''}
\item \textbf{Transcript excerpt:} ``Subtype number two, mixed ASD with developmental delay. Making up about 19\% of autistic people in the study fell into this group.''
\item \textbf{Verdicts:} Gemini-3: \textsc{unsupported}; GPT5: \textsc{overstated}; Grok: \textsc{unverifiable}
\item \textbf{Reasoning (Gemini-3 Flash):} The transcript explicitly identifies the four subtypes and their percentages as 37\%, 19\%, 34\%, and 10\%, stating that the group with developmental delays and life skill challenges (Subtype 2) accounts for 19\%.
\end{itemize}

\vspace{6pt}
\noindent\textbf{Example 5 --- Concept mixup}
\begin{itemize}
\setlength{\itemsep}{1pt}
\setlength{\parskip}{0pt}
\item \textbf{Video:} Nail-Patella syndrome (uploaded 2023-03-19)
\item \textbf{Claim:} \textit{``These nail abnormalities most commonly affect the thumbnails, with the severity decreasing with each finger.''}
\item \textbf{Transcript excerpt:} ``The most consistent clinical manifestation involves nail changes like absent, underdeveloped, split, ridged, or pitted nails, which most commonly affect the thumbnails, with the changes decreasing in frequency with each finger.''
\item \textbf{Verdicts:} Gemini-3: \textsc{overstated}; GPT5: \textsc{unverifiable}; Grok: \textsc{supported}
\item \textbf{Reasoning (Gemini-3 Flash):} The transcript states that the frequency of nail abnormalities decreases from the thumb to the other fingers, whereas the claim characterizes this gradient as a decrease in severity.
\end{itemize}

\end{document}